\documentclass[%
reprint,
groupedaddress,
 amsmath,amssymb,
 aip,
 apl,
floatfix,
]{revtex4-1}

\usepackage{xcolor} %for colored text in comments

\usepackage{graphicx}% Include figure files
\usepackage{dcolumn}% Align table columns on decimal point
\usepackage{bm}% bold math
\usepackage[mathlines]{lineno}% Enable numbering of text and 
\usepackage[version=4]{mhchem} %for chemical names
\usepackage[colorlinks=true,allcolors=blue]{hyperref} %hyperrefs

\begin{document}

%\preprint{APS/123-QED}

\title{A perspective on hybrid quantum opto- and electromechanical systems}%\thanks{A footnote to the article title}%

\author{Yiwen Chu}
\email{yiwen.chu@phys.ethz.ch}
\affiliation{Department of Physics, ETH Z\"urich, 8093 Z\"urich, Switzerland}

\author{Simon Gr\"oblacher}
\email{s.groeblacher@tudelft.nl}
\affiliation{Kavli Institute of Nanoscience, Department of Quantum Nanoscience, Delft University of Technology, 2628CJ Delft, The Netherlands}

\date{\today}

\begin{abstract}
Quantum opto- and electromechanical systems interface mechanical motion with the electromagnetic modes of optical resonators and microwave circuits. The capabilities and promise of these hybrid devices have been showcased through a variety of recent experimental advances that demonstrated exquisite control over the quantum state of solid-state mechanical objects. In this perspective, we offer an overview of the current state, as well as an outlook of the future directions, challenges, and opportunities for this growing field of research. We focus in particular on the prospects for ground state cooling of mechanical modes and their use in quantum circuits, transducers, and networks.
\end{abstract}

\maketitle

One of the first model systems we encounter when learning about quantum mechanics is a mass on a spring, described as a quantum harmonic oscillator. The concept of phonons, the quanta of mechanical motion, is also used to explain a wide variety of phenomena ranging from the BCS theory of superconductivity~\cite{BCS1957} to the scattering of photons in optical materials~\cite{Shen1965}. Being able to actually observe, control, and make use of the quantum behavior of solid-state mechanical objects, however, has proven to be very challenging. Over the past decade, the development of high quality mechanical modes and the ability to engineer their interactions with the environment has led to steady progress in the fields of optomechanics and electromechanics and enabled proof-of-principle experiments using hybrid systems to manipulate mechanical objects in the quantum regime. In this perspective, we briefly summarize these developments in this new field of quantum acoustics, the quantum behavior of mechanical motion in a solid-state object, and focus on a forward-looking view of the challenges and opportunities for future opto- and electromechanical devices.

\begin{figure}[ht]
\centerline{
\includegraphics[width=0.5\textwidth]{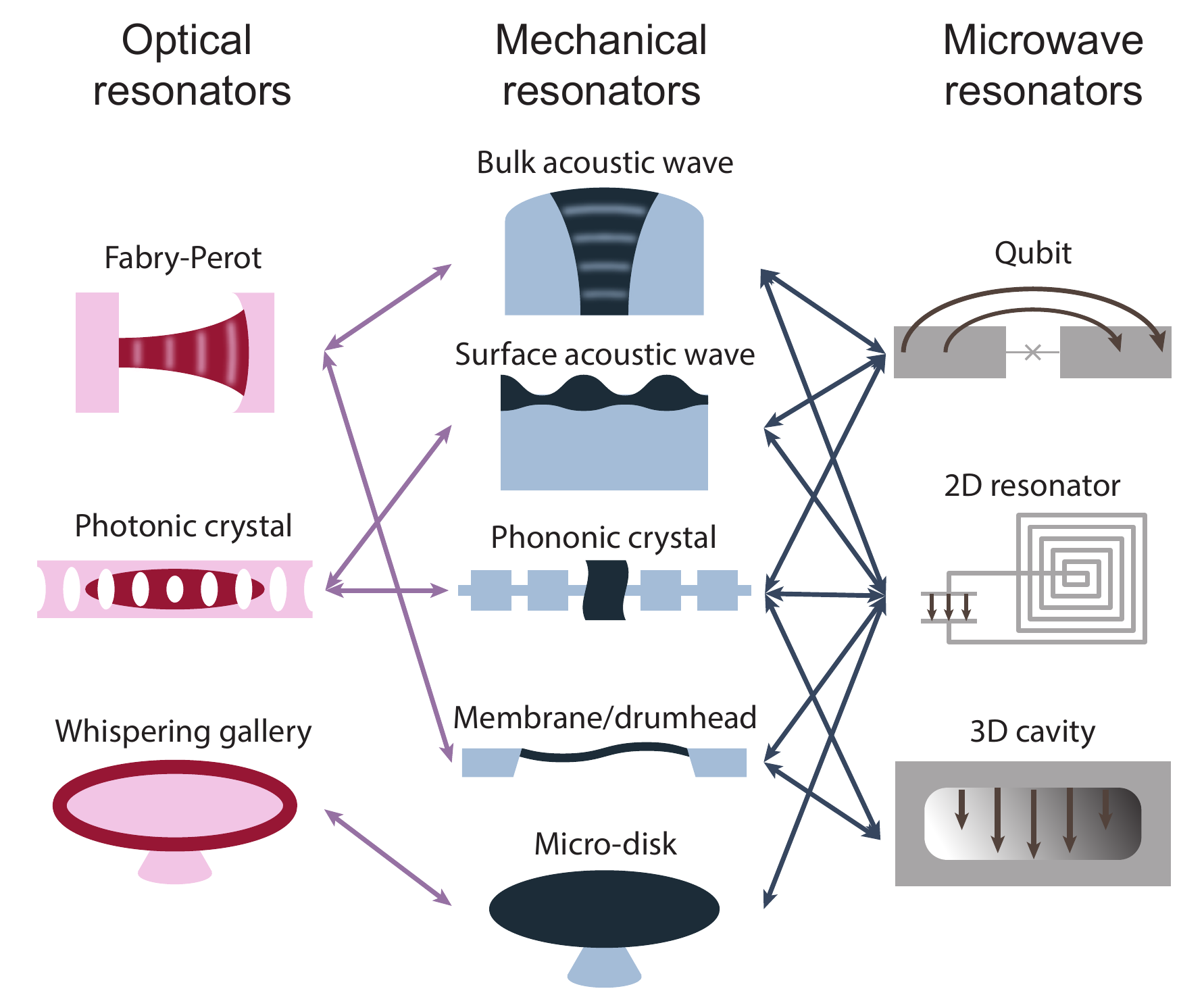}}
\caption{Different types of optical, mechanical, and microwave resonators that have been used in experimental demonstrations discussed in this perspective. Arrows indicate demonstrated couplings between systems.}
\label{fig1}
\end{figure}

The motional degree of freedom of atomic systems such as trapped ions~\cite{Molmer1999} and ultracold atomic ensembles~\cite{Murch2008} has been extensively studied and utilized for many years. The nascent field of quantum acoustics focuses instead on the mechanical motion of solid state objects composed of a much larger number of atoms. These objects range from membranes, beams, and phononic crystals, to levitated nano-objects, to resonators for bulk and surface acoustic waves (see Figure~\ref{fig1}). Similarly to atomic systems, these massive objects need to be well isolated from their environments and their mechanical modes need to be cooled such that their quantum behavior is not overwhelmed by thermal motion. Unlike atoms, however, the mechanical modes of these objects do not naturally couple to easily accessible quantum mechanical degrees of freedom and are typically well described by classical harmonic oscillators. In order to overcome these limitations, opto- and electromechanical systems are engineered to provide an electromagnetic (EM) "control knob" for mechanical resonators in the form of light and electrical circuits. These electromagnetic components provide crucial access to quantum mechanical ingredients such as non-linearities for creating non-Gaussian states and a low thermal occupation environment for cooling. Solid-state mechanical objects can also be coupled to, for example, spin qubits, quantum dots, and atoms to create a large variety of other hybrid systems, but we will not focus on these approaches here and would instead like refer to several excellent review articles~\cite{Kurizki2015,Lee2017,Clerk2020}.

The rest of this article will be organized into four topics, each one relating to a potential application of quantum opto- and electromechanical systems. We begin with the cooling of mechanical modes to the quantum ground state, which is the basic starting point for quantum control of mechanical motion. We will then consider the prospects of using mechanical resonators as new types of quantum circuit elements in the subfield of circuit quantum \textit{acousto}-dynamics (circuit QAD). Finally, we will discuss the potentially crucial role that mechanical resonators can play in long-distance quantum networks, both as memories that can be entangled through optomechanical interactions with photons, and as transducers that can convert quantum information between the microwave (MW) and optical domains.

\section{Ground state cooling}

Of particular importance for quantum experiments with mechanical resonators is the amount of classical noise that is coupled into the mechanical mode of interest from the surrounding thermal environment. The thermal occupation of the oscillator directly defines the suitability of the device for quantum experiments and determines the thermal decoherence rate of any quantum state. It is therefore highly-desirable to initialize the mechanical mode in or close to its quantum ground state, which has been one of the biggest challenges in the early days of the fields of opto- and electromechanics. Various approaches have been pursued to successfully realize mechanical resonators in the ground state, with the most prominent being either working directly with high-frequency modes that can be cryogenically pre-cooled, in analogy to SC systems, or by using radiation-pressure cooling to remove unwanted thermal excitations of the mechanical mode through the coupled near-zero temperature light field. Being able to use a variety of cooling techniques also enables the exploitation one of the major attractions of mechanical oscillators -- the fact that they are engineered quantum systems. This allows them to not only rely on naturally occurring resonances, but rather makes them a versatile platform to design and tailor a device to the specific needs of the experiment. Typical parameters include the mechanical linewidth (or Q factor), coupling strength, as well as the frequency, which usually ranges anywhere from the Hz to the GHz regime. These unique properties make them ideal building blocks for hybrid quantum devices. In this section, we would like to briefly review the two main cooling approaches that are currently pursued in experiments, along with their advantages and remaining challenges.

\begin{figure}[ht]
\centerline{
\includegraphics[width=0.5\textwidth]{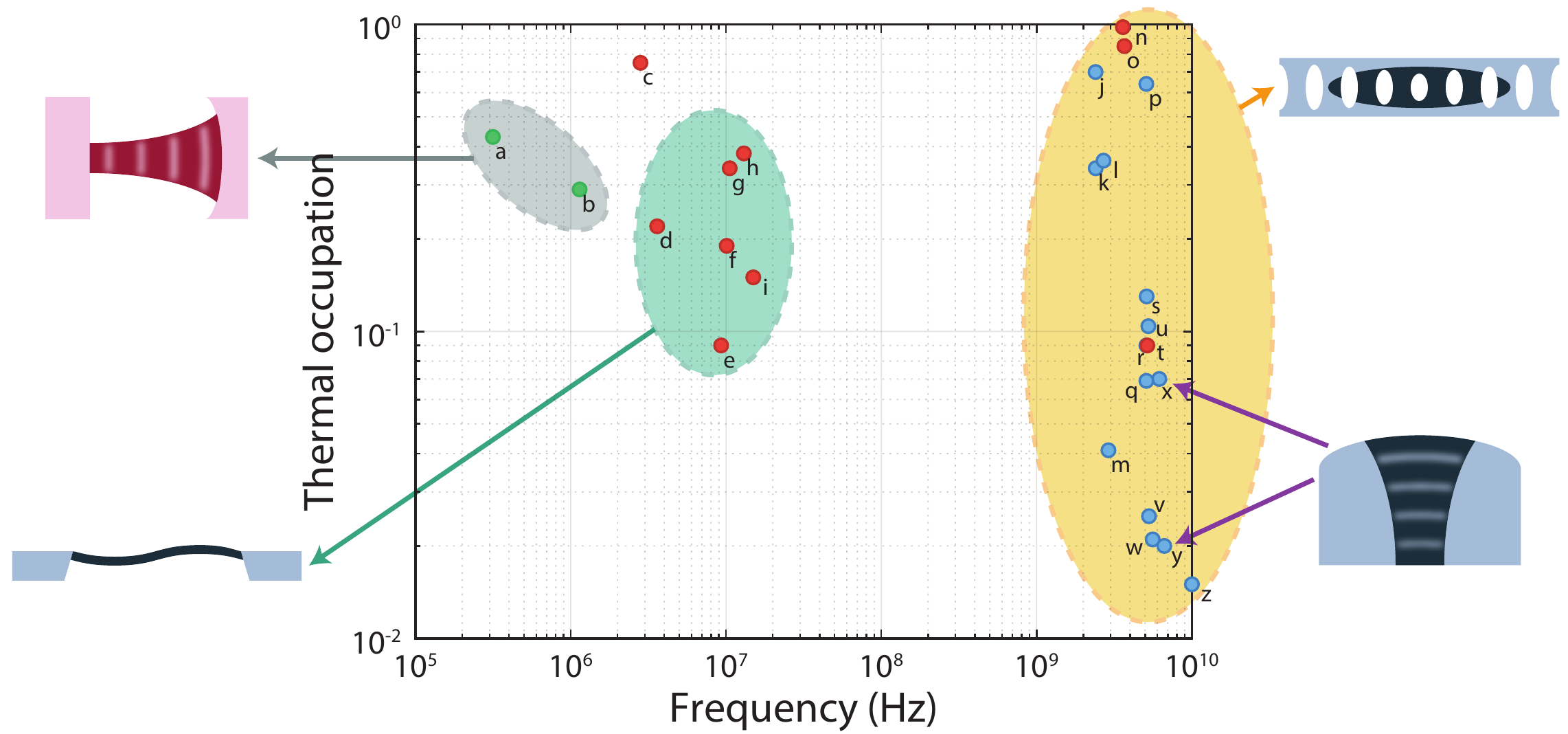}}
\caption{Comparison between various experiments demonstrating ground state cooling of the mechanical mode ($n<1$). Plotted are the smallest achieved thermal occupation vs.\ the mechanical resonator's frequency. The different cooling techniques are color-coded as follows:\ green (active feedback), red (cavity sideband cooling) and blue (cryogenic). The various types of systems can be roughly divided into the highlighted regions, with GHz phononic crystals on the right (orange) with the exception of BAW systems, superconducting drums in the center (dark green) and optical cavity based experiments top left (gray). References:\ a~\cite{Delic2020}, b~\cite{Rossi2018}, c~\cite{Barzanjeh2019}, d~\cite{Wollman2015}, e~\cite{Reed2017}, f~\cite{Clark2017}, g~\cite{Teufel2011b}, h~\cite{Pirkkalainen2015}, i~\cite{Lecocq2015}, j~\cite{Ramp2019}, k~\cite{Arrangoiz-Arriola2019}, l~\cite{Forsch2020}, m~\cite{Stockill2019}, n~\cite{Meenehan2014}, o~\cite{Chan2011}, p~\cite{Mirhosseini2020}, q~\cite{Riedinger2018}, r~\cite{Marinkovic2018}, s~\cite{Wallucks2020}, t~\cite{Qiu2020}, u~\cite{Hong2017}, v~\cite{Riedinger2016}, w~\cite{Meenehan2015}, x~\cite{OConnell2010}, y~\cite{Chu2017}, z~\cite{Ren2019}. }
\label{fig2}
\end{figure}

High-frequency mechanical modes, typically around a few GHz, can in principle be directly cooled into their quantum ground state using a mK thermal environment  inside a dilution refrigerator (see Figure~\ref{fig2}). While this limits the possible device parameters, it has enabled the realization of several seminal quantum experiments with mechanical oscillators, both in the microwave~\cite{OConnell2010,Chu2018,Satzinger2018,Arrangoiz-Arriola2019} and the optical regime~\cite{Riedinger2016,Riedinger2018,Marinkovic2018}. Besides the limitation to GHz modes, heating through finite thermalization owed to the very small thermal conductivity at mK temperatures, as well as through optical absorption, poses some remaining challenges and restrictions~\cite{Meenehan2015,Hong2017,Hauer2018}. Novel approaches with optimized thermal anchoring~\cite{Safavi-Naeini2014,Ren2019} will lead to devices that will allow for practical quantum technologies using mechanical systems.

Working at lower frequencies requires additional cooling of the mechanical modes, as significant residual thermal energy remains even at mK temperatures. Various optical cooling approaches, such as sideband- and active-feedback cooling~\cite{Aspelmeyer2014} have however resulted in ground state cooled mechanical oscillators. This was achieved through both cryogenic pre-cooling for a reduced initial occupation~\cite{Teufel2011b,Chan2011,Rossi2018} and, most recently, starting from room temperature~\cite{Delic2020}. As these experiments remain highly challenging by themselves, there have been very few experiments using radiation-pressure cooled devices to demonstrate true quantum behavior, such as nonclassical states of motion or entanglement~\cite{Wollman2015,Pirkkalainen2015,Lecocq2015,Reed2017,Ockeloen-Korppi2018}, all with electromechanical systems. With further improvements in coupling strengths, mechanical quality factors, as well as absorption mitigation and handling, such experiments will however soon become more routine and allow for quantum experiments at only moderate or even ambient temperatures.

\section{Circuit QAD}

Circuit QAD makes use of hybrid systems involving a mechanical resonator coupled to a non-linear, MW frequency superconducting (SC) qubit. In close analogy to the well-established field of circuit quantum electro-dynamics (cQED), the coupling between the mechanical mode and the SC qubit is a bilinear interaction, which we distinguish from the fundamentally nonlinear optomechanical interaction~\cite{Krause2015b,Hauer2019} used in experiments involving mechanical modes coupled to linear MW resonators. Coupling strengths in the range of 100's of kHz to 10's of MHz can be achieved by placing a piezoelectric material in the electric field of the qubit mode~\cite{OConnell2010,Chu2018,Satzinger2018,Arrangoiz-Arriola2019} or by engineering an appropriate voltage bias on a qubit capacitor plate that also supports a mechanical mode~\cite{Viennot2018}. Over the last few years, devices based on these principles have been used to create quantum states of mechanical motion, perform full quantum tomography of these states~\cite{Chu2018,Satzinger2018}, and spectroscopically distinguish the number of phonons in a mechanical mode \cite{Viennot2018,Arrangoiz-Arriola2019,Sletten2019}.

These demonstrations show that circuit QAD is situated at an exciting juncture, where the large variety of tools that have been developed in circuit QED for creating, manipulating, and measuring quantum states in EM systems are now within reach for mechanical systems. This, however, begs the question of what makes mechanical resonators uniquely useful or interesting as quantum circuit elements and what their advantages and drawbacks are compared to MW resonators. 
An obvious challenge in circuit QAD is the need to combine experimental techniques that have previously been carefully optimized individually for SC circuits, mechanical resonators, and piezoelectric transducers. Generally speaking, increasing the coupling between different elements also tends to further compromise their isolation from lossy environments. For example, SC circuits are usually made on extremely low-loss dielectric substrates such as sapphire and silicon. While these materials can also support long-lived mechanical modes, they are not piezoelectric. In fact, piezoelectricity and a low loss tangent may be fundamentally conflicting requirements due to the possibility of phonon radiation as an additional source of MW loss~\cite{Scigliuzzo2020}. Indeed, even BAW-based circuit QAD devices, which have the highest coherences demonstrated so far, exhibit qubit lifetimes ($\sim$7~$\mu$s) and mechanical quality factors ($\sim$3 million)~\cite{Chu2018} that fall short of state-of-the-art values~\cite{Paik2011,Renninger2018}. 
% , even in devices that prioritized simplicity of fabrication and maintaining quantum coherence over maximizing coupling strengths. 
However, given that this is a relatively new field, there is still a large parameter space to be explored, and it is likely that drastic improvements can be made as long as the potential trade-off between coupling strengths and coherence properties is taken into account when developing new materials, designs, and fabrication methods.

On the other hand, mechanical resonators can actually provide a new way of increasing the coherence times of information stored in a quantum circuit. Two important recent developments have laid the foundations for potentially using mechanical resonators as long-lived quantum memory elements in a SC quantum computer. The first is an alternative strategy for encoding and performing error correction on quantum information in circuit QED by storing it in the harmonic oscillator modes of a MW resonator rather than the two level systems of SC qubits~\cite{Terhal2020}. The second is the realization of ultra-high quality factor mechanical resonators at GHz frequencies, which can be near-resonantly coupled to SC qubits. For example, the longest lived GHz MW resonators used in cQED~\cite{Reagor2016} have Q's on the order of $10^7-10^8$, a value that has also been measured in bulk acoustic wave (BAW) resonators~\cite{Renninger2018} and far exceeded by phononic crystal (PC) resonators~\cite{Tsaturyan2017,MacCabe2019} with Q's of more than $10^{10}$. We note that, for quantum memory applications, the coherence time including dephasing is a more important figure of merit, and linewidths for BAW resonators ($\sim$300~Hz) indicate comparable coherence to $T_2$ values of MW resonators measured using Ramsey-type measurements ($\sim$0.7~ms)~\cite{Renninger2018,Reagor2016}. Furthermore, since mechanical resonators and EM resonators are susceptible to different loss mechanisms, it remains to be seen which system will provide an easier path toward even longer lived quantum memories. If these mechanical resonators can be incorporated into SC circuits without degrading their coherence properties, they can provide a promising alternative for quantum information storage. 
 
Another unique advantage that mechanical resonators offer as quantum circuit elements is their compact size. Since the speed of sound is 4-5 orders of magnitude slower than the speed of light in most materials, a mechanical resonator will be smaller by roughly the same factor compared to an EM resonator with the same fundamental frequency. This can already be observed in several existing circuit QAD devices, where a single qubit is used to address many mechanical modes localized to a much smaller volume than the SC circuit~\cite{Arrangoiz-Arriola2019,Sletten2019,Chu2018}. These modes can either be higher-order modes of a single BAW or surface acoustic wave (SAW) resonator~\cite{Sletten2019,Chu2018} or belong to physically distinct PC resonators~\cite{Arrangoiz-Arriola2019}. The compactness of these modes also implies better isolation and smaller crosstalk between devices. Several architectures for quantum memories and processors have already been proposed to take advantage of this unique property of circuit QAD systems~\cite{Pechal2019,Hann2019}. Experimentally, each type of mechanical resonator offers different advantages and challenges in the context of multimode operation. For example, while BAW and SAW resonators provide a wealth of modes without the need for fabricating many physical devices, it is difficult to individually control each mode's frequency and coupling strength to the qubit, and schemes need to be developed so that they can be selectively addressed in quantum operations.

\section{Quantum transduction}

\begin{figure}[ht]
\centerline{
\includegraphics[width=0.5\textwidth]{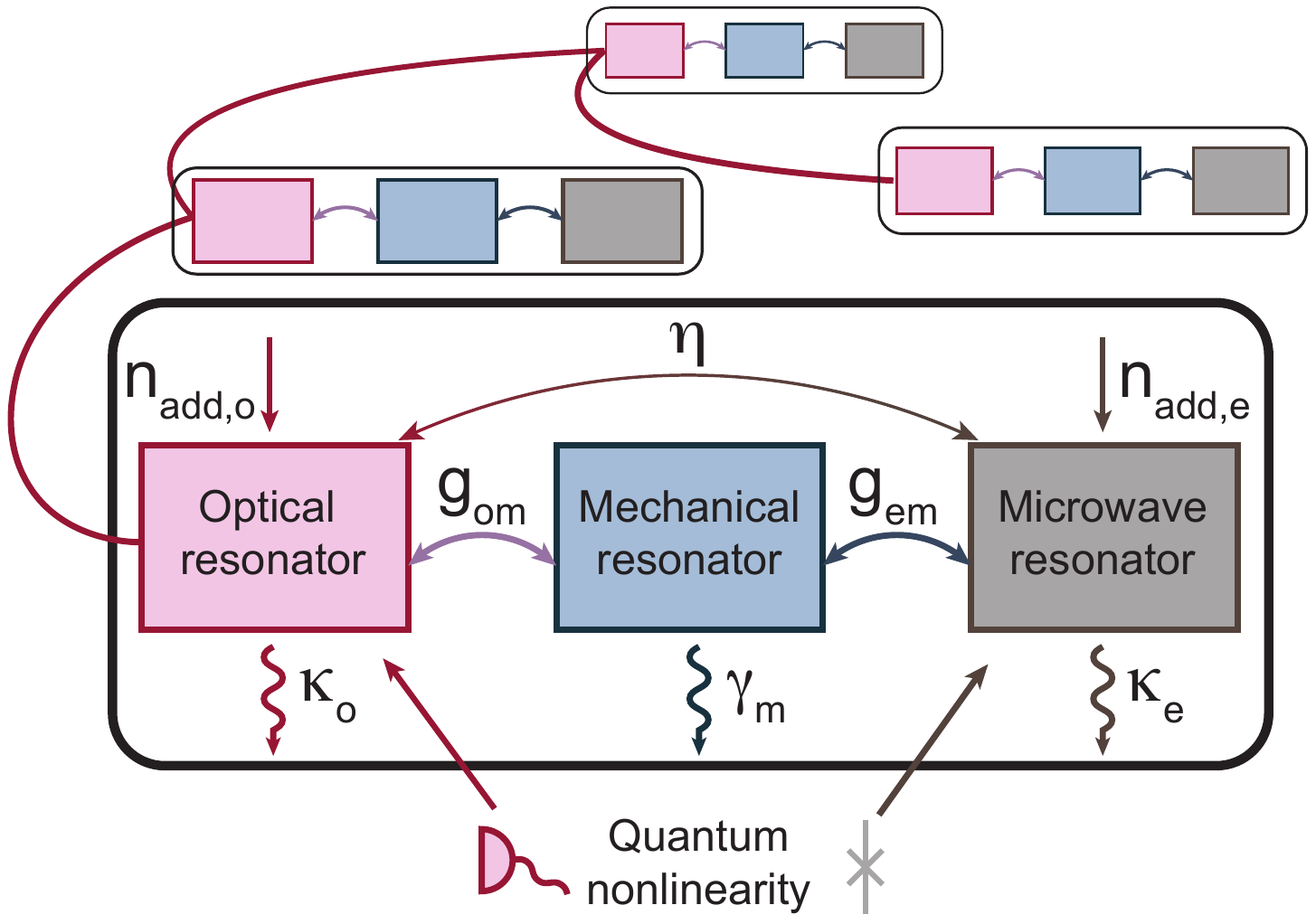}}
\caption{Ingredients for quantum transduction between the microwave and optical domains, using a mechanical resonator as intermediary. The optomechanical and electromechanical coupling strengths are $g_\text{om}$ and $g_\text{em}$, respectively. The mechanical resonator has a dissipation rate $\gamma_\text{m}$. The optical and microwave dissipation rates $\kappa_\text{o}$ and $\kappa_\text{e}$ can include both internal losses and coupling to input/output ports. The overall transduction process has an efficiency $\eta$, while the added noise referred to the optical and microwave inputs are $n_\text{add,o}$ and $n_\text{add,e}$, respectively. Also shown are a single photon detector and a Josephson junction, which are sources of quantum nonlinearity mentioned in the text.}
\label{fig3}
\end{figure}

Our discussion so far makes it clear that mechanical resonators can be efficiently interfaced with both MW frequency circuits and optical light in the quantum regime, particularly in the telecom band, which makes them a promising way to create a link between these two EM systems. Achieving this link would unite one of the leading platforms for quantum computing with the most convenient carrier of quantum information over large distances, making it a crucial component in building a network of SC quantum processors~\cite{Lambert2020,Lauk2020}. A desirable specification for such a quantum transducer is the ability to deterministically convert quantum states between the MW and optical domains while preserving their quantum properties. Achieving this goal entails meeting three basic requirements~\cite{Zeuthen2020}, as illustrated in Figure~\ref{fig3}:\ First, the transducer should be able to directly interface with nonlinear quantum resources, such as the non-linearity of a SC Josephson junction or a single-photon detector. Second, the transducer should have high efficiency so there is a low probability of losing even a single quantum of information during the transduction process, which generally requires a high cooperativity for both the electro- and optomechanical processes~\cite{Hill2012}. Third, the transducer should introduce much less than one quantum of added noise~\cite{Wu2020,Zeuthen2020}. While there are many ways of quantifying the ``quantumness" of the transduction process, failure to satisfy any of these requirements would preclude, for example, measuring quantum statistics or negativity in the Wigner function of the transduced state. While recent work has demonstrated relatively high efficiency transduction of Gaussian states~\cite{Higginbotham2018} or transducers incorporating SC qubits~\cite{Mirhosseini2020}, satisfying all three requirements remains an outstanding challenge.

We now discuss several mechanical platforms for quantum transduction. The highest transduction efficiency of 47$\%$ for classical signals was demonstrated using a silicon nitride membrane coupled to a niobium superconducting resonator and a Fabry-P\'{e}rot optical cavity~\cite{Higginbotham2018}. Such a system is able to simultaneously reach matched electromechanical and optomechanical cooperativities of $\sim$60, so that the efficiency is mostly determined by limitations in the external coupling and optical mode matching. Achieving low added noise with the MHz frequency mechanical modes of these membranes continues to be a challenge, although transduction with tens of added noise photons have been demonstrated by incorporating laser cooling and classical feed-forward. Another unaddressed experimental challenge in such low-frequency systems is the incorporation of qubits or other sources of non-classical states. 

Using high-frequency phononic crystal resonators allows the transducer to benefit directly from operating the mechanical mode close to the quantum ground state, minimizing the added noise in the transduction process. Similar to cQAD, sometimes conflicting requirements for material properties requires careful material choice and parameter optimization. These include strong piezoelectricity, low microwave and low mechanical loss, large optomechanical coupling, and minimal optical absorption. Two main approaches to realizing such a high-frequency transducer exist, where the \textit{homogeneous} one is performed with the transducer fabricated fully from a piezo-electric material, including the optomechanical cavity. This has been done with \ce{AlN}~\cite{Vainsencher2016}, several III-V materials, such as \ce{GaAs}~\cite{Balram2016} and \ce{GaP}~\cite{Stockill2019,Schneider2019}, as well as \ce{LiNbO3}~\cite{Jiang2020}. The other approach is based on a hybrid system, where the optomechanical part is fabricated from \ce{Si} for example, coupling to the microwave input through an added piezoelectric resonator, such as \ce{AlN}~\cite{Han2020,Mirhosseini2020} or \ce{LiNbO3}~\cite{Jiang2019,Shao2019} or directly through capacitive electro-mechanical transduction~\cite{Arnold2020}. This \textit{heterogeneous} approach combines the advantages of both systems, directly benefiting from the excellent performance of \ce{Si} as an optomechanical quantum system, while allowing for relatively large efficiencies for the electro-mechanical coupling. This approach has not only allowed for low-added noise transduction, but recently even for coupling to a superconducting qubit~\cite{Mirhosseini2020}. Some of the major remaining challenges include complex fabrication procedures for both homo- and heterogeneous approaches, such as unreliable etching techniques and combining high quality films in the desired parts of the device without affecting the supporting material quality and absorption, respectively, as well as separating optical fields and superconducting materials from one another. Further improvements in reliably designing triple-resonant devices~\cite{Ramp2020} (MW - piezo - mechanics) or including in-situ tuning mechanisms will also have to be demonstrated, in addition to operating with a large enough bandwidth.

Systems where surface acoustic waves (SAWs) act as an intermediary have also received significant attention. SAWs allow the use of technologies for creating mechanical quantum states using microwaves and inter-digitated transducers (IDT) and then coupling them to photonic crystal cavities. This has lead to low-added noise transduction~\cite{Forsch2020}, with significant challenges in the efficiencies of coupling the SAWs to the resonant mechanical modes remaining. This is in big part owed to the size mismatch between a typical surface acoustic wave and a mechanical resonator, as well as practical limitations of reflected waves at various interfaces. Creative ideas such as focusing IDTs~\cite{Vainsencher2016} have been tested to mitigate these challenges, introducing additional problems such as polarization mixing. Optimizing a hybrid system incorporating SAW and integrated piezoelectric devices will probably allow us to avoid and overcome some of these current limitations.

BAW resonators have shown high cooperativity coupling to both SC qubits~\cite{Chu2018} and to infrared light through Brillouin interactions~\cite{Kharel2018b}. Although circuit QAD experiments have confirmed that these GHz frequency acoustic modes are in the quantum ground state at dilution refrigerator temperatures, ground state operation has yet to be demonstrated in a cavity optomechanics scenario, which would be a natural next step toward integration into a low-noise quantum transducer. The large volume to surface ratio of BAW resonators may offer a distinct advantage in mitigating the detrimental effects of laser heating on both the mechanical mode and SC circuit. However, this relatively new platform for quantum electro- and optomechanics still poses many open questions and has a largely unexplored parameter space. For example, finding an appropriate material for the BAW resonator will likely be a compromise between not only minimizing MW, acoustic, and optical dissipation properties, but also achieving sufficiently high photoelastic constants and convenient Brillouin frequencies for infrared light. 
We would also like to highlight that there are several alternative approaches to the quantum transduction challenge that do not involve mechanical resonators. These make use of, for example, magnons, rare-earth ions, as well as direct electro-optic coupling~\cite{Lambert2020,Lauk2020,Clerk2020}. In particular, the latter approach has a long history in the classical domain as the mechanism behind electro-optical modulators (EOM) and is now intensely pursued to also operate in the quantum regime, with low-added noise, large bandwidth and efficient transduction. While the extremely large bandwidth is one of the main advantages over the mechanical systems, low-noise operation still remains an outstanding challenge~\cite{Bagci2014,Wang2018,Fan2018,Jiang2020}.

\section{Optomechanical quantum networks}

While quantum transduction has been one of the most promising directions of using mechanical systems in the quantum regime, the experimental realization of quantum entanglement between mechanical resonators also demonstrates the possibility to integrate them into future quantum networks~\cite{Kimble2008} (cf.\ Fig.~\ref{fig3}). Such a network would allow to connect various distinct quantum resources over long distance quantum links. Again, some of the highly attractive features of using mechanical oscillators for direct quantum information processing tasks can be directly attributed to the ease and versatility of coupling them to other quantum systems, as well as their high degree of engineerabilty, allowing them to operate natively in the telecom band for example. The first step in demonstrating this potential were recent experiments on creating entangled states between several mechanical resonators~\cite{Riedinger2018,Ockeloen-Korppi2018}, as well as between mechanics and optical fields~\cite{Palomaki2013,Marinkovic2018,Barzanjeh2019,Chen2020}. Entanglement is one of the basic resources required for any quantum network, and both discrete as well as continuous variable entanglement has been shown using mechanical systems. While much potential for improvements remains, in particular in terms of the achievable entanglement rate, state fidelity, as well as general efficiency, recent progress not only shows that it is in principle possible to create mechanical entangled states, but that such states can already be created over tens of meters and used as resources to violate a Bell-type inequality~\cite{Marinkovic2018}. This has significant implications on the usability and security of the entanglement for a quantum network.

In order to realize mechanical systems that can not only act as simple transducers but potentially even as quantum repeaters in a hybrid quantum network architecture, demonstrating entanglement is however not sufficient. In particular, interfacing the optical states with a quantum memory is another key requirement. Ideally, the mechanics itself can directly act as such a memory device, as has recently been shown~\cite{Wallucks2020}. Additionally, as a further step in the proof-of-principle demonstrations of mechanics-based repeater architectures, a quantum teleportation protocol has to be realized. Theoretical proposals for such protocols have been put forward in both the continuous variable ~\cite{Mancini2003,Felicetti2017} and discrete variable regimes~\cite{Pautrel2020,Li2020}. Realizing such experiments seems within reach of several state-of-the-art mechanical systems and will be the next milestone on the path to mechanics-based quantum network devices.

\section{Conclusions}

While much of the focus of optomechanics over the past years has been on proof-of-principle experiments, the rapid development of the field has also clearly shown that mechanical systems have significant potential for becoming part of real quantum technologies. As highlighted in the various potential applications discussed in this perspective, some of the advantages of using mechanical oscillators can be directly attributed to the relative ease with which they can be coupled to other quantum systems and engineered to have a diverse range of properties. This versatility will allow for mechanical objects to function as quantum transducers not only between EM systems at microwave and optical frequencies, but also between defect centers~\cite{Lee2017}, rare-earth ions~\cite{Molmer2016,Wang2020}, atoms~\cite{Karg2020}, ions~\cite{Schuetz2015}, etc. In addition to serving as an interface between other quantum systems, mechanical modes themselves offer the promise of more coherent and hardware-efficient memories or processors in a quantum network. The last ten years have seen immense progress in fundamental experiments using opto- and electromechanical systems. With most of the basic quantum effects now experimentally demonstrated, we envision the next ten years to become the decade of real mechanical quantum applications. Thanks to their unique versatility, we foresee that these systems will evolve into indispensable quantum technologies.

\begin{acknowledgments}
We would like to thank Andreas Wallucks, Bas Hensen, Hugo Doeleman, John Davis, Kartik Srinivasan, and Robert Stockill for valuable discussions. This work is supported by the Foundation for Fundamental Research on Matter (FOM) Projectruimte grant (16PR1054), the European Research Council (ERC StG Strong-Q, 676842), and by the Netherlands Organization for Scientific Research (NWO/OCW), as part of the Frontiers of Nanoscience program, as well as through Vidi (680-47-541/994) and Vrij Programma (680-92-18-04) grants.
\end{acknowledgments}

\vspace{.5cm}

\noindent The source data for the figures is available at \href{https://doi.org/10.5281/zenodo.3932217}{10.5281/zenodo.3932217}.

\bibliography{Mirror}

\end{document}